\documentstyle[prl,aps]{revtex}
\draft
\begin{document}
\input{epsf}
\twocolumn[\hsize\textwidth\columnwidth\hsize\csname@twocolumnfalse\endcsname
\title{Complex Dynamics of Piecewise Linear Model of the Van~der~Pol Oscillator\\
Under External Periodic Force}
\author{E.S.~Mchedlova and L.V.~Krasichkov}
\address{Department of Nonlinear Processes, Saratov State University \\
83 Astrakhanskaya, Saratov 410012, Russia}

\date{\today}
\maketitle
\begin{abstract}
The electronic model of Van~der~Pol oscillator with piecewise linear
$V-I$--characteristics of nonlinear element is proposed. It is carried out
the experimental investigations of behaviour of the Van~der~Pol 
oscillator under external periodic force. 
The parameter plane "the amplitude of external
force -- the frequency of external force" for the dynamical regimes  of the
oscillator is experimentally plotted. It is shown in experiments and numerical
simulations that transition to chaos in the oscillator takes place through
the period doubling cascade. It is revealed that the internal structure of
the synchronization tongues has the characteristic view as for 
{\it the crossroad area}.

\end{abstract}

\narrowtext
\vskip1pc]

The Van~der~Pol oscillator is a classical example of self-oscillatory system.
More than seventy years after its appearing, the oscillator has become the
main model self-oscillatory system with two dimensional phase space
\cite{r1,r2,r3}.
The classical experimental setup of the system is the oscillator with vacuum
triode \cite{r2,r3}. The mathematical model for the system is well known second
order ordinary differential equation with cubic nonlinearity -- the
Van~der~Pol equation.
The Van~der~Pol equation is basic model for oscillatory processes 
in physics, electronics, biology, neurology,
sociology and economics \cite{r3a}.

It was much attention dedicated to investigations of  
the peculiarities of the Van~der~Pol
oscillator behaviour under external periodic (sinusoidal) force and, in
particular, the synchronization phenomena and the dynamical chaos appearing
(see e.g., \cite{r3b,r4,r5,r6,r7}).

The investigations of the forced Van~der~Pol oscillator behaviour have
carried out by many researchers. So, in 1927 Van~der~Pol and Van~der~Mark
\cite{r1} in their investigations of the oscillator behaviour 
in the relaxation
oscillation regime found that the subharmonical oscillations are appeared
during changes of natural frequency of the system. Moreover, the authors
noted appearing of "irregular noise" before transition from one
subharmonical regime to another. It seems to be it was one of the first
observations of chaotic oscillations in the electronic circuit. Later,
in 1945, Cartwright and Litlewood \cite{r8} during their  
analyzing the Van~der~Pol equation
with large value of nonlinearity parameter have shown 
that the singular solutions exist. 
In 1949 Levinson \cite{r9}, analytically analyzing the Van~der~Pol equation,
substitutes the cubic nonlinearity for piecewise linear version and shows
that the equation has singular solutions in the case also.
It was revealed  that chaotic behaviour is appeared in the Van~der~Pol 
equation with smooth nonlinearity by period
doubling bifurcations \cite{r7} and crossroad area 
is observed in synchronization tongues \cite{r9a}.

However, up to the present day the difficulties connected with creation of
the electronic model of the nonautonomous Van~der~Pol oscillator 
which demonstrates transition to chaos in physical experiment and
will allows adequate numerical simulation are remained. The efficient enough
way for creation of such kind models is building of the systems with 
piecewise linear voltage--current characteristics \cite{r10,r11}. 
In the paper \cite{r11} on the basis of numerical simulation shown
that in piecewise linear model of sinusoidally forced Van~der~Pol oscillator
the transition from periodic oscillations to chaotic ones is possible. The
transition takes place through cascade of period doublings.

In this paper the results of experiment and numerical
simulation of the sinosoidally forced 
Van~der~Pol oscillator model with symmetric
piecewise linear voltage-current characteristic of nonlinear element are
discussed.

{\it Experimental circuit and models.} The
classical dimensionless 
form the nonautonomous Van~der~Pol oscillator is described by
equation
\begin{eqnarray}
\label{eq1}
\ddot {x} - \varepsilon (1 - x^{2})\dot {x} + x = a\cos \omega t,
\end{eqnarray}
where $x$ is the state variable, $\varepsilon$ is the nonlinearity
parameter, $a$ and $\omega$ are the amplitude and frequency of external
force, respectively. 

\begin{figure}
\begin{center}
\leavevmode
\hbox{%
\epsfxsize=6.5cm
\epsffile{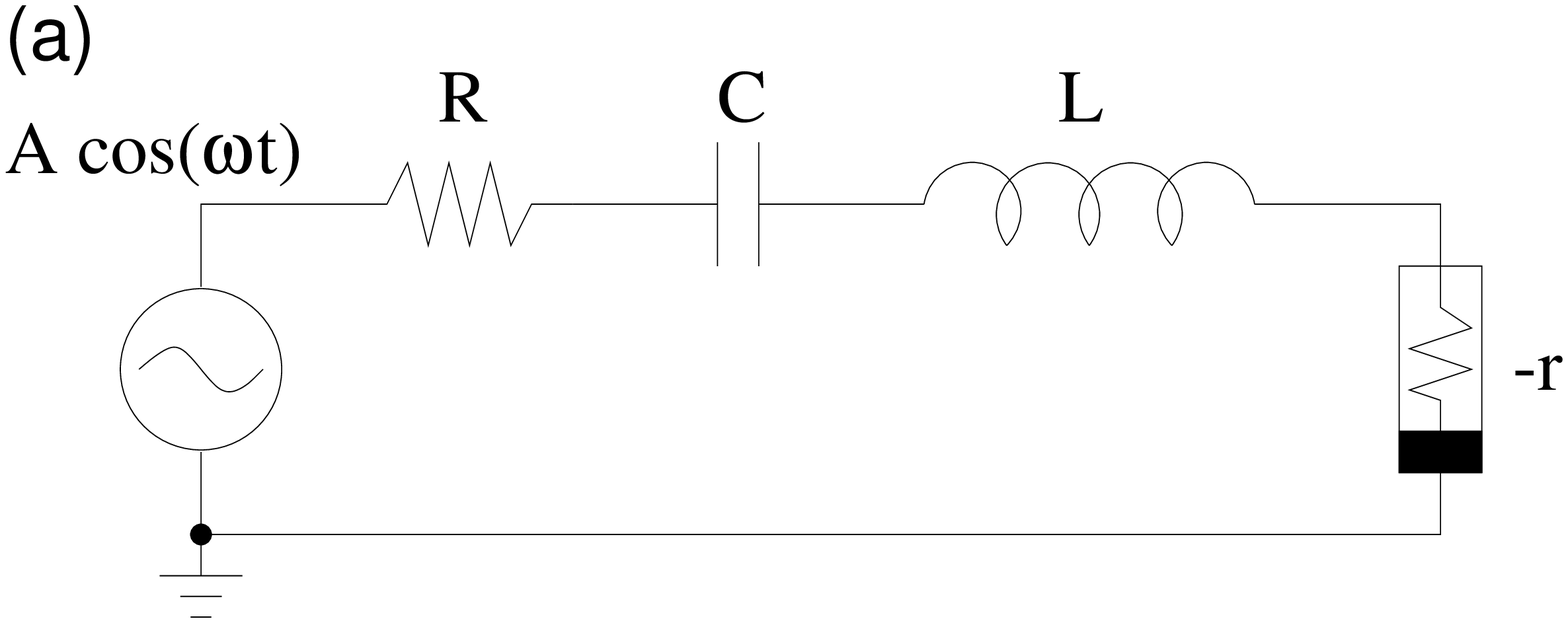}}
\hbox{%
\epsfxsize=3.3cm
\epsffile{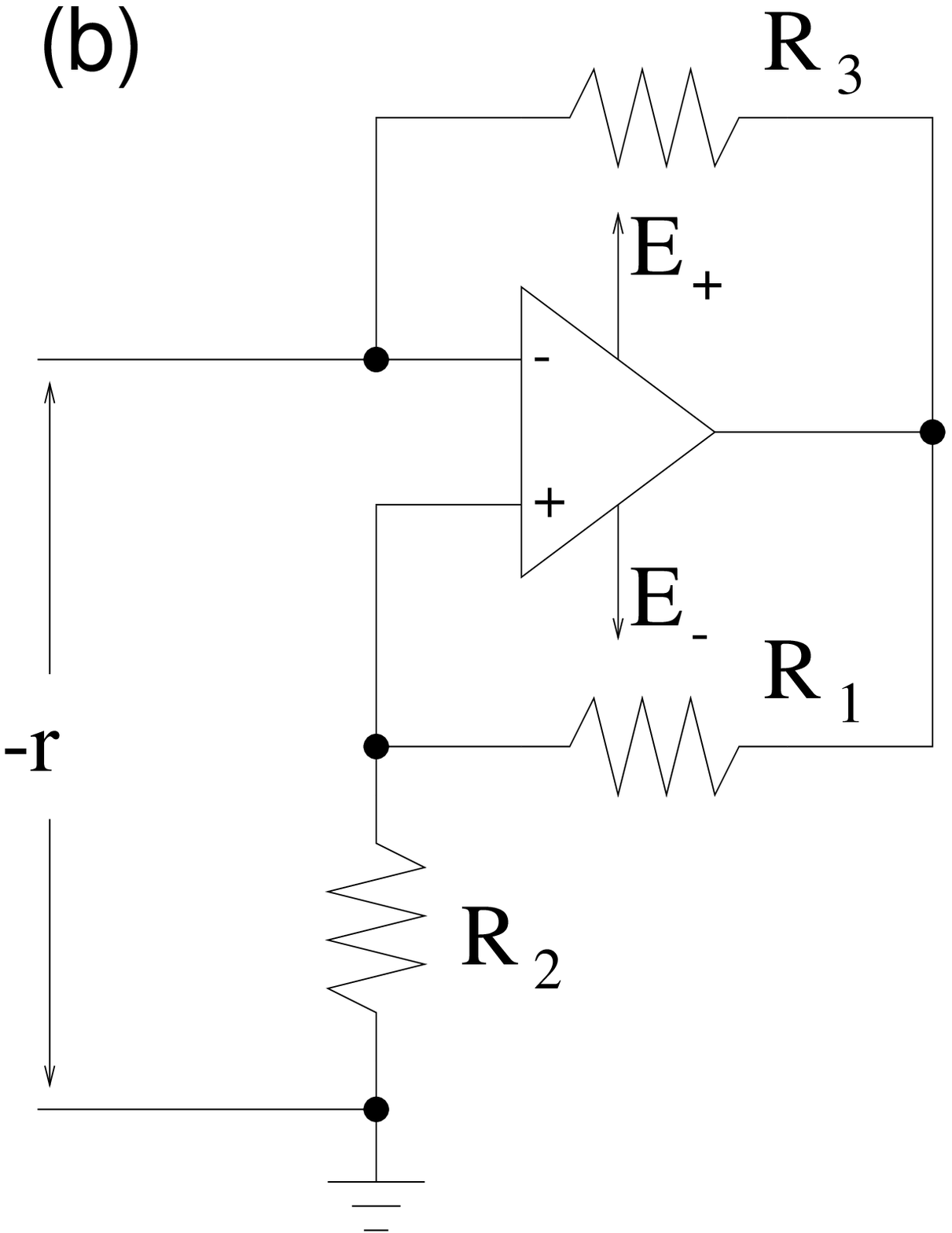}}
\end{center}
\caption{Schematic diagram of the Van~der~Pol oscillator under external
sinusoidal force (a) and the schematic diagram of nonlinear element with
piecewise-linear characteristic (b).}
\label{fig1}
\end{figure}

\begin{figure}
\begin{center}
\leavevmode
\hbox{%
\epsfxsize=6.5cm
\epsffile{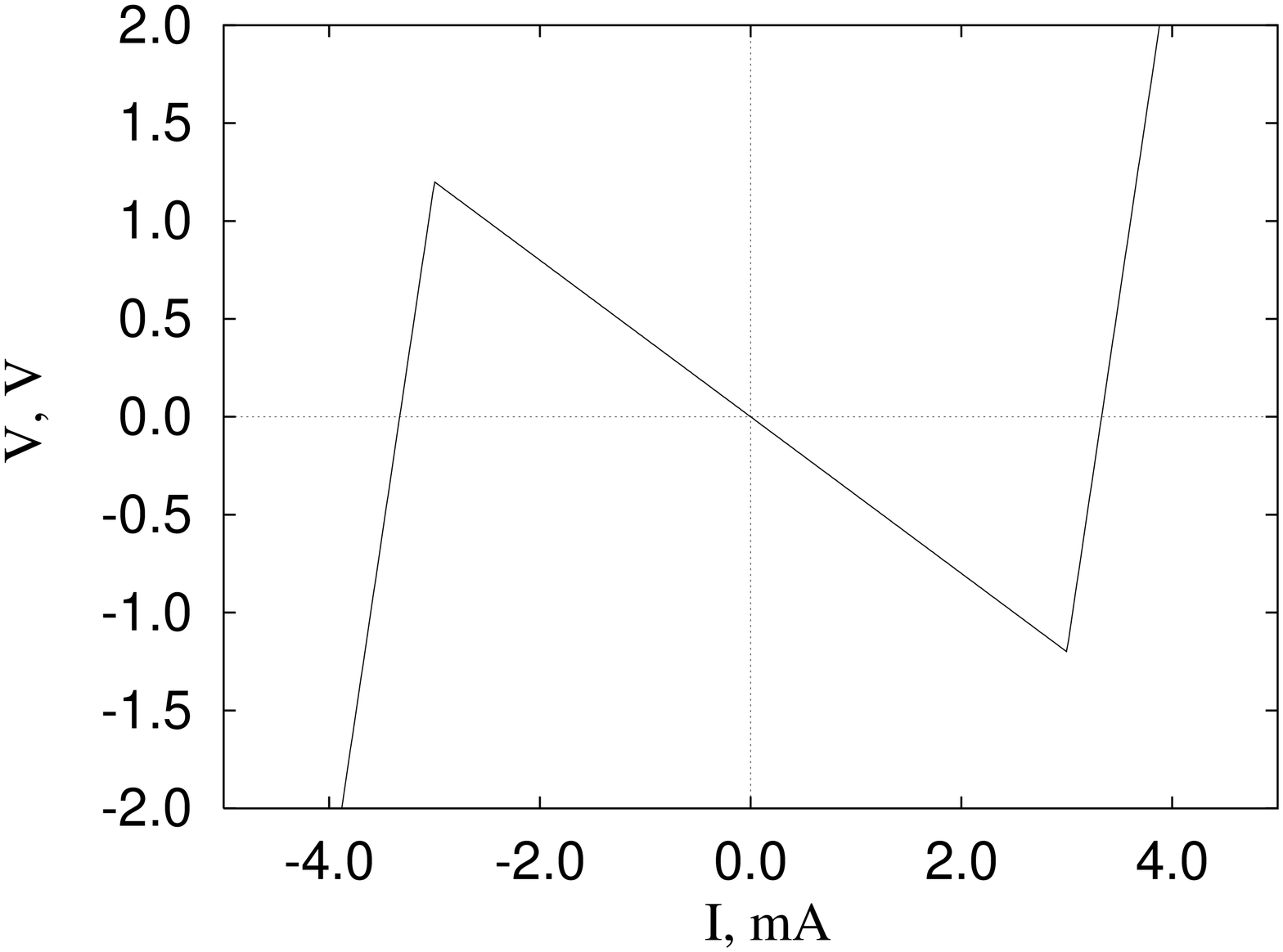}}
\end{center}
\caption{The piecewise-linear voltage-current characteristic of nonlinear
element which depicted in Fig.~\ref{fig1}b.}
\label{fig2}
\end{figure}

For building of electronic circuit of the Van~der~Pol oscillator which
would be adequate for modeling by (\ref{eq1}) the serial circuit could be
used (Fig.~\ref{fig1}a).  The
nonlinear element $( - r)$ in the circuit (Fig.~\ref{fig1}a) must obey with
symmetrical characteristics with negative resistance region. The nonlinear
resistor with three-segment piecewise linear $S$--type voltage-current
characteristic is created on an operation amplifier 
\cite{r13}. 
The scheme is shown in Fig.~\ref{fig1}b, 
where the rest parameters are R$_1$=R$_3$=3.6~k$\Omega$, R$_2$=400~$\Omega$,
$E_{+} = 12$~V, $E_{-} = - 12$~V. The voltage-current characteristic
measured in experiment for the above mentioned parameter values is presented
in Fig.~\ref{fig2}. The resistance at negative region of nonlinear element
voltage-current characteristic is $R_{-}=-400$~$\Omega$ and at positive
region is $R_{+}=3600$~$\Omega$.

On the basis of Kirhgoff laws the equations for the Van~der~Pol oscillator
model (Fig.~\ref{fig1}a) could be written as
\begin{eqnarray}
\label{eq2}
 L{\frac{{dI}}{{dt}}} &+& V_{C} + RI + V(I) = A\cos \omega t, \nonumber \\
 {\frac{{dV_{C}}} {{dt}}} &=& {\frac{{I}}{{C}}}, 
\end{eqnarray}
where $V_{L} $ is the voltage at the inductor, $V_{R} $ is the voltage at
the resistor, $V_{C} $ is the voltage at the capacitor, $V(I)$ is the
function describing piecewise linear voltage--current
characteristic of the nonlinear element "$ - r$", 
$A$ and $\omega $ are the amplitude and
frequency of external force, respectively. Introducing new dimensionless
quantities: 
${\tau = \omega _{0} t}$, $x = I / I_{0}$, 
$y = V / V_{0}$, $f(x) = V(I) / V_{0}$, 
${\Omega = \omega / \omega _{0}}$, 
$\omega _{0} = 1 / \sqrt {LC}$, 
$\varepsilon = \left( V_{0} / I_{0} \right) \sqrt { C / L } $,
$A_{0} = \left( A / I_{0} \right) \sqrt { C / L } $, 
$G = R \sqrt{ C / L} $ in term of which (\ref{eq2}) become
\begin{eqnarray}
\label{eq3}
 \dot {x} &=& - Gx - \varepsilon y - \varepsilon f(x) + A_{0} \cos \Omega \tau
, \nonumber \\
 \dot {y} &=& {\frac{{1}}{{\varepsilon}} }x,
\end{eqnarray}
where $I_{0} $ and $V_{0} $ are the values of current and voltage in
extremum of the voltage-current characteristic (Fig.~\ref{fig2}), 
$\left( { \cdot} \right) = {d / {d\tau}}$.

Along with the piecewise linear form of $f(x)$ 
in the equations (\ref{eq3}) the arbitrary function $f(x)$ could be used. 
So, if nonlinearity in eqs.~(\ref{eq3}) is governed by 
${f(x) = {\frac{{1}}{{3}}}x^{3} - x}$
then additive "$ - Gx$" could be removed by rescaling and  
finally the eq.~(\ref{eq1}) is obtained. 
Thus, the experimental scheme (Fig.~\ref{fig1}a) and the appropriate
equation (\ref{eq3}) are adequately described the classical 
Van~der~Pol oscillator
under external force (\ref{eq1}).

{\it Experimental results.} For systems under external force, as a
rule, the
synchronization phenomena of the natural oscillations by external periodical
force are investigated 
at the parameter plane "the amplitude -- the
frequency of external force". Most frequently the synchronization regimes
are marked by (1:1), (1:2), (1:3) and so on. 
Such notations of the synchronization
regimes $(m:n)$ show the relation of the natural frequency of gerenator's
oscillations ($m$) to the frequency of the external force ($n$). 
There are so called main synchronization regimes marked 
by (1:1), (1:2), (1:3), (3:1), (2:1) etc.
Between such regimes the more complicated multiloop regimes of high
order are placed. All these regimes have the rational relation of
frequencies. Moreover, the regions (quasiperiodical oscillations) with
irrational relation of frequencies exist.

Experimental investigation of the driven Van~der~Pol oscillator behaviour is
in plotting of synchronization regimes at the parameter plane. For
experiments in this work the following parameters were used:
$R$=25.8~$\Omega $, $C$=83.5~nF, $L$=4.37~mH (Fig.~\ref{fig1}a). 
The resistance $R$ is total
resistance (excluding $-r$, $R_L$=23.3~$\Omega$ is 
the resistance of inductor) 
of the circuit. The boundaries of oscillatory regimes were
defined from the phase portrait view during the changes of controlling
parameters. In the some cases the changing of both parameters simultaneously
is needed because the oscillator demonstrates the multistable behaviour. The
multistability is appeared as existence of the histeresis boundaries for
different oscillatory regimes. It could be proposed there are the regions
which could be reached for the both parameter changing only.

When the parameter plane has been plotted the
synchronization regimes were identified by phase portraits. 
The different regimes depicted as
the synchronization tongues. Generally speaking, 
the tongues of higher order are placed under
the tongues of the lower order, for example, the synchronization tongue
(1:2) could be partially placed under the synchronization tongue (1:1). 
The regions
with period doubling is marked with subscript letters (e.g., $(1:2)_{2} $,
$(1:2)_{4}$ ), the regions with chaotic behaviour is marked with subscript
$C$ (e.g., (1:2)$_{C})$.

\begin{figure}
\begin{center}
\leavevmode
\hbox{%
\epsfxsize=\columnwidth
\epsffile{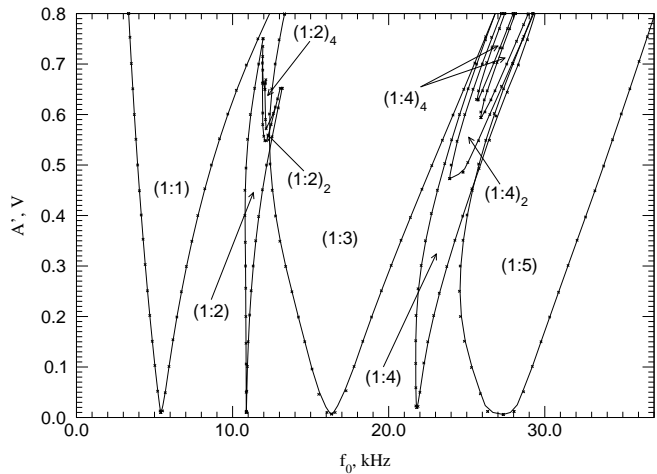}}
\end{center}
\caption{The parameter plane "the amplitude of external force -- the
frequency of external force" ($A=\sqrt{2}\approx1.414~A'$, $\omega=2 \pi f_0$) 
for the Van~der~Pol oscillator 
(Fig.~\ref{fig1}) with piecewise-linear 
$V$--$I$--characteristicÓ of nonlinear element (experiment).}
\label{fig3}
\end{figure}

\begin{figure}
\begin{center}
\leavevmode
\hbox{%
\epsfxsize=4.3cm
\epsffile{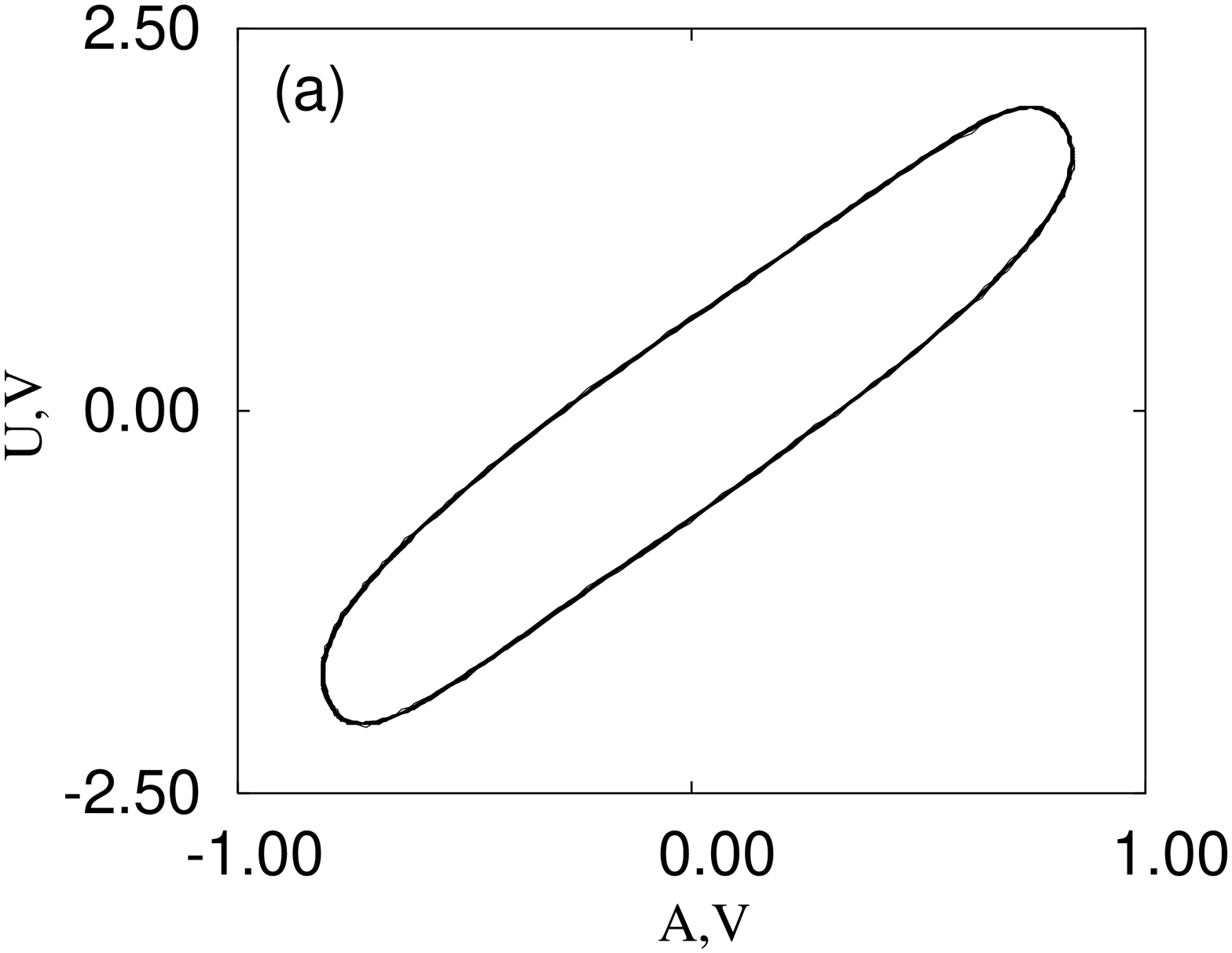}
\epsfxsize=4.3cm
\epsffile{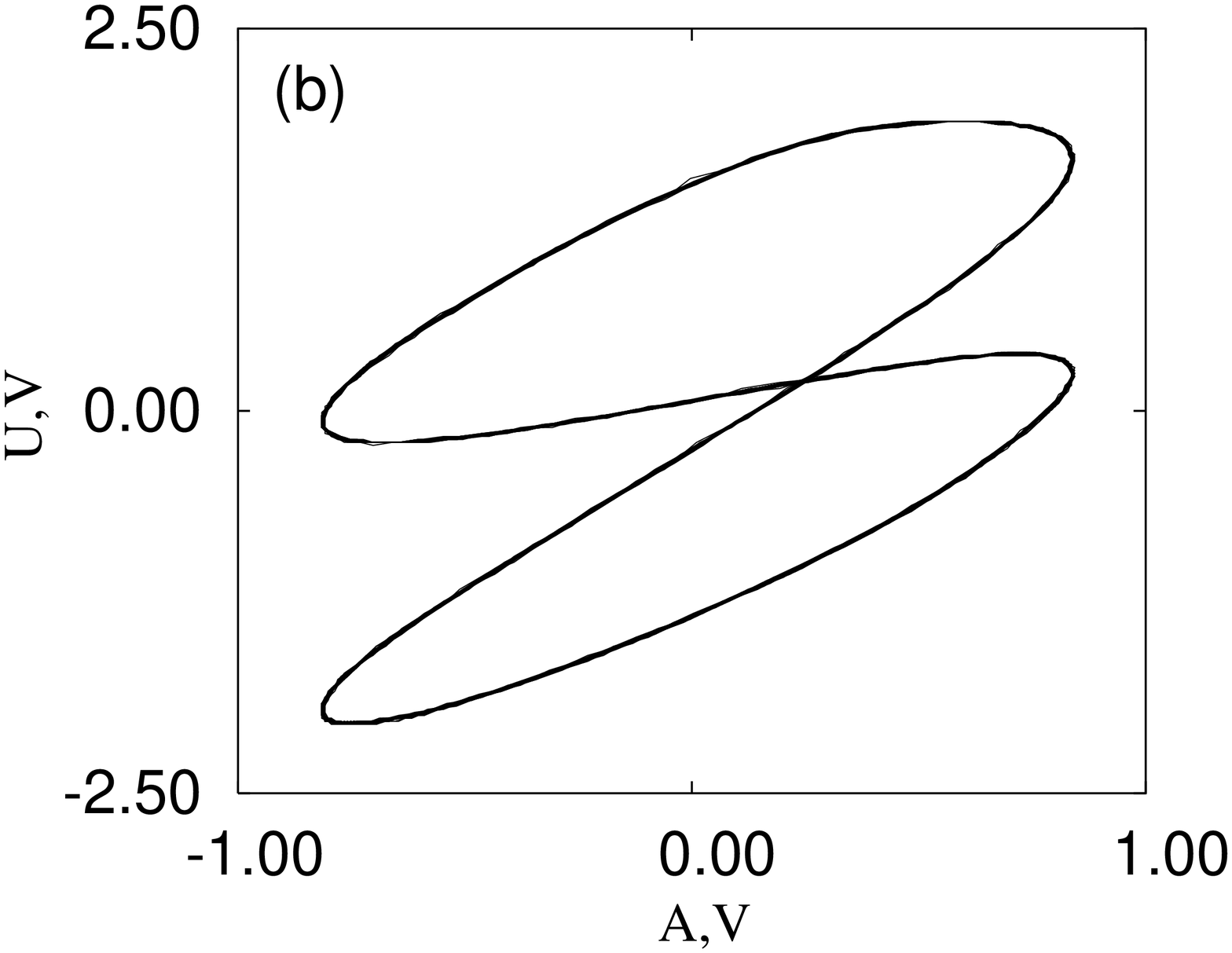}}
\hbox{%
\epsfxsize=4.3cm
\epsffile{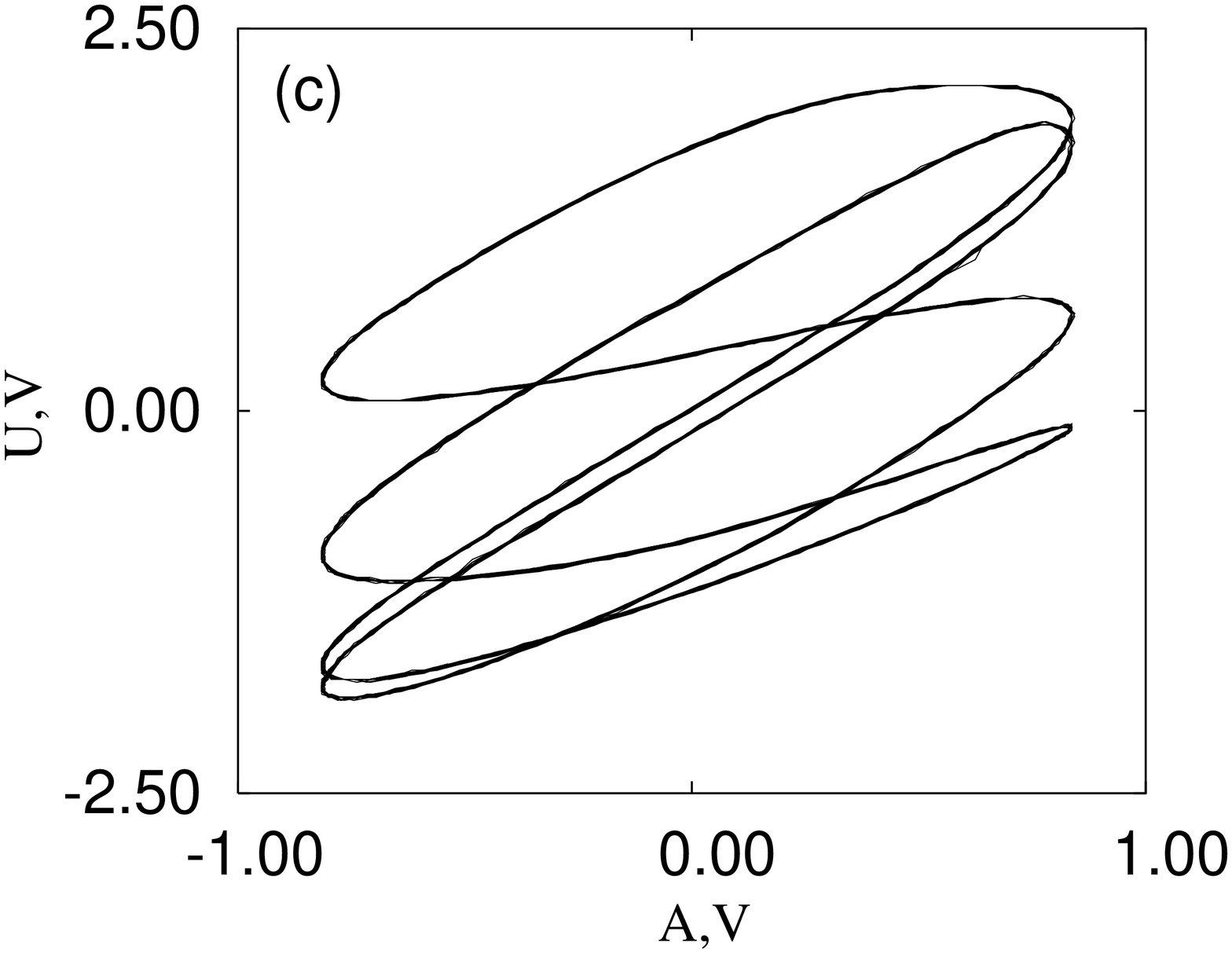}
\epsfxsize=4.3cm
\epsffile{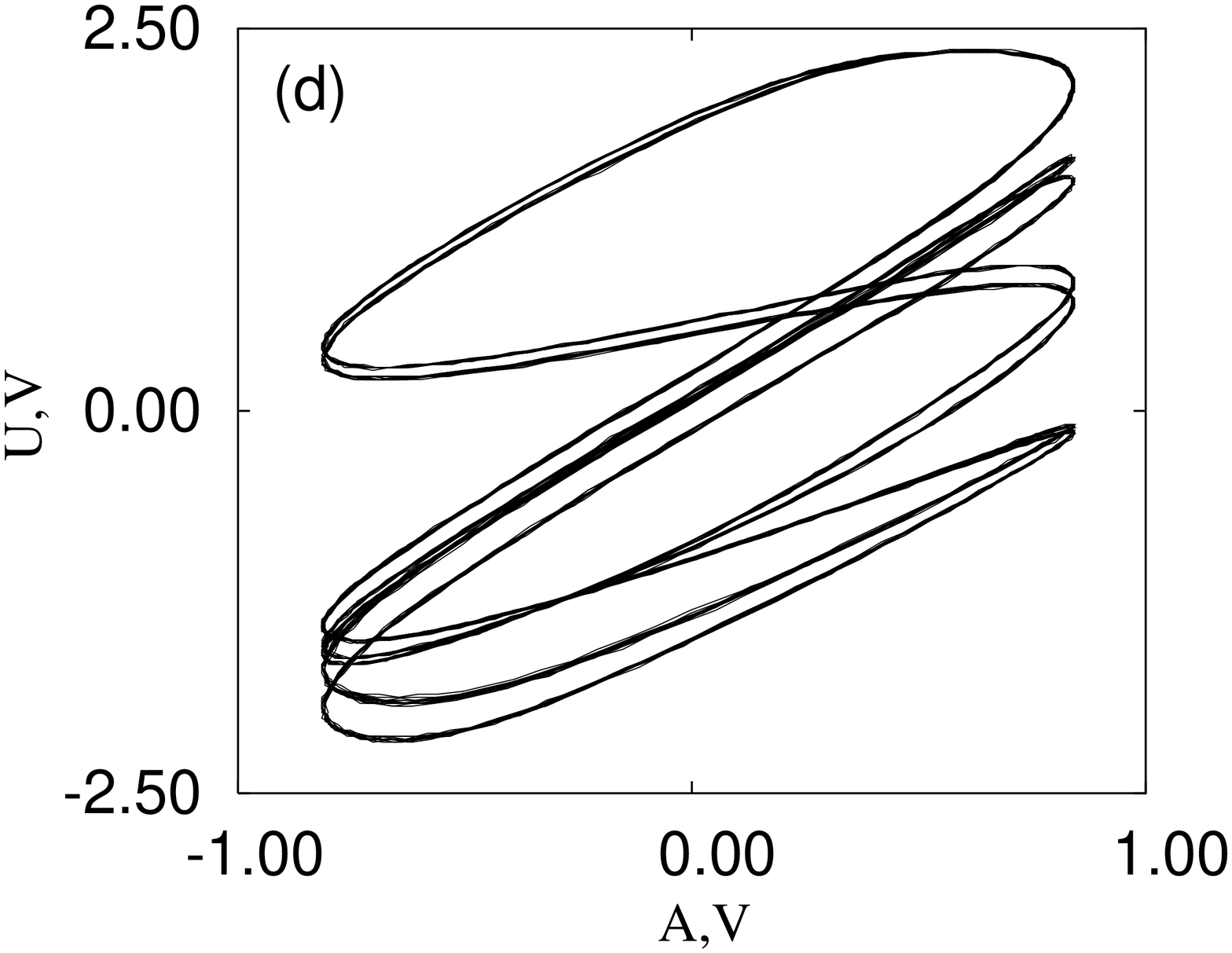}}
\hbox{%
\epsfxsize=4.3cm
\epsffile{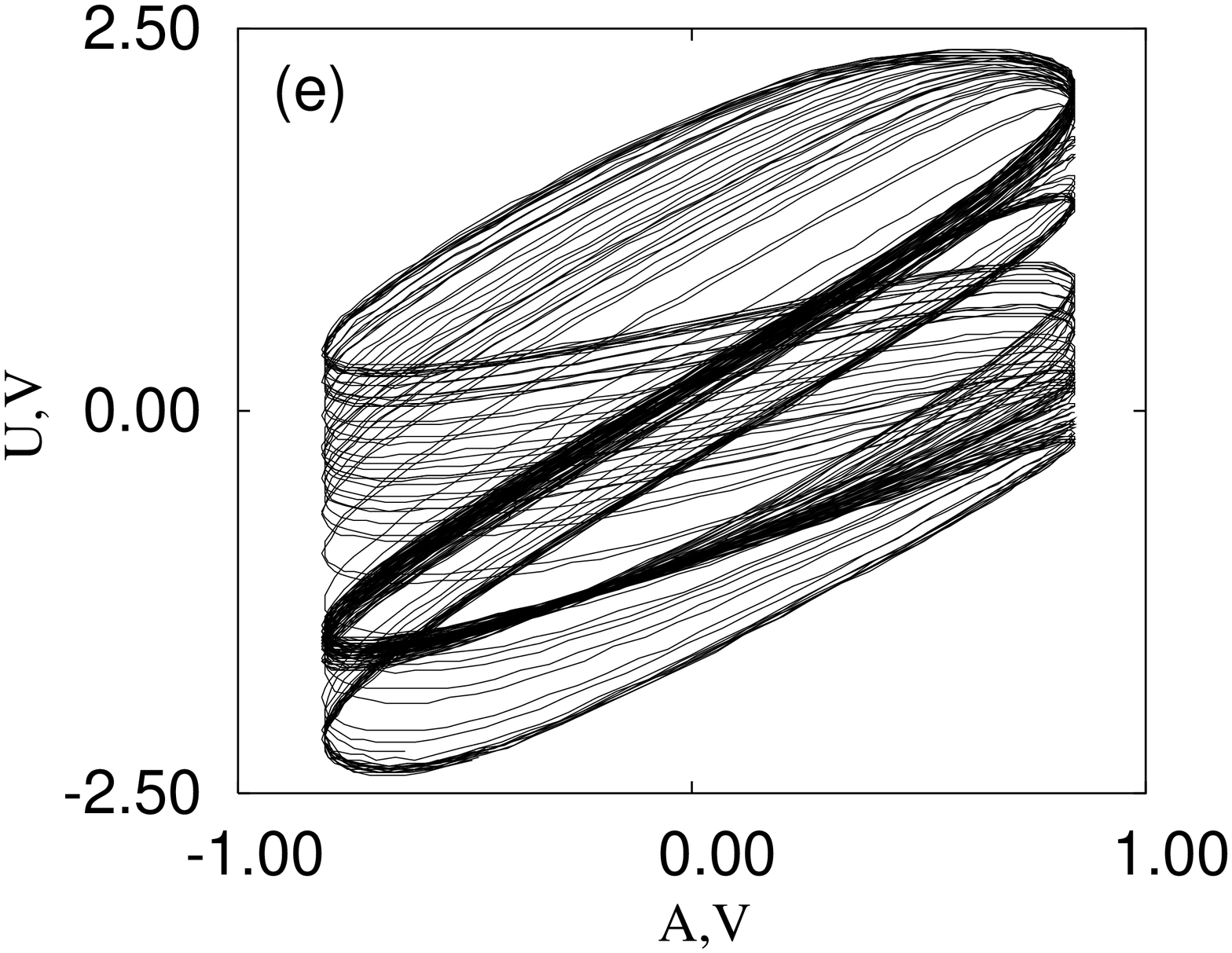}
\epsfxsize=4.3cm
\epsffile{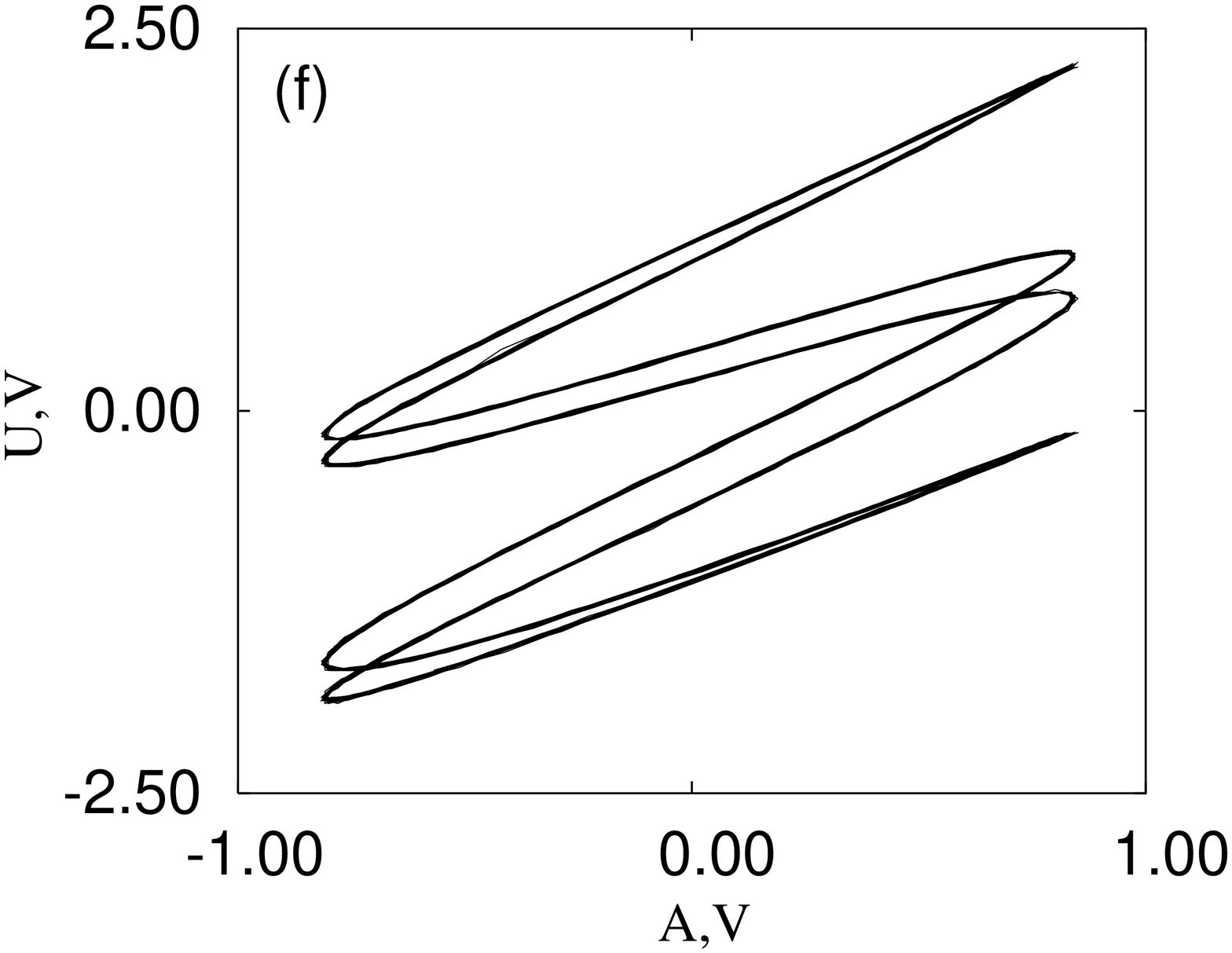}}
\end{center}
\caption{The phase portraits for $A'$=0.60~V and $f_{0}=6.5$~kHz (1:1)
(a), $f_{0}$ = 11.7~kHz (1:2) (b), $f_{0}$ = 12.0~kHz $(1:2)_{2}$ (c), 
$f_{0}$ = 12.15~kHz $(1:2)_{4}$ (d), $f_{0}$ = 12.25~kHz (1:2)$_{C}$ (e), 
$f_{0}$ = 24.7~kHz (1:4) (f) 
for the characteristic dynamical regimes depicted at the
parameter plane (Fig.~\ref{fig3}).}
\label{fig4}
\end{figure}

The parameter plane "the amplitude -- the frequency of
external force" for the Van~der~Pol oscillator (Fig.~\ref{fig1}) with the
piecewise-linear $V$--$I$--characteristics of nonlinearity 
(Fig.~\ref{fig2}) is shown in Fig.~\ref{fig3}.
The main synchronization tongues are shown in Fig.~\ref{fig3} only, 
although, many narrow tongues with rational
frequencies relation are observed in experiment.
The phase portraits for typical dynamical regimes depicted at
the parameter plane (Fig.~\ref{fig3}) are shown in Fig.~\ref{fig4}. 
In Fig.~\ref{fig4} the amplitude of
external force changes for the horizontal axis and the voltage in the point
of connection of capacitor $C$ and inductor $L$ -- for vertical axis. The
data were sampled with an analog to digital converter with 8--bit resolution
at sampling rate $2\cdot10^7$ samples per second, the length of time series
is 3000 points. 

As it is seen in Fig.~\ref{fig3} the overlapping of the 
synchronization tongues (1:2)
and (1:3) in the vicinity $f_0\sim12.6$~kHz, $A'\sim0.6$~V takes place. The
chaotic regimes with strange attractor could be seen in the synchronization
tongues (1:2) and (1:4), more precisely in $(1:2)_{4}$, $(1:4)_{4}$ regions. 
It is
easily seen that the internal structure of the synchronization tongue (1:4)
has the view that is characteristic for the bifurcation composition called
{\it crossroad area} \cite{r14}.

\begin{figure}
\begin{center}
\leavevmode
\hbox{%
\epsfxsize=6.5cm
\epsffile{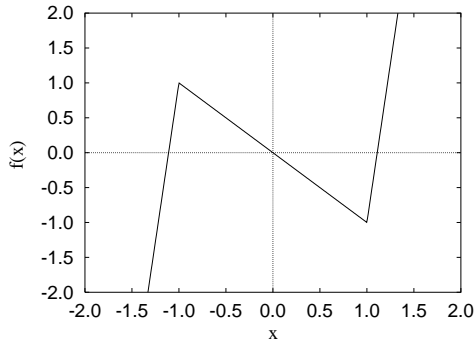}}
\end{center}
\caption{The rescaled piecewise-linear voltage-current characteristics of the
nonlinear element (Fig.~\ref{fig2}), described by eq.(\ref{eq4}).}
\label{fig5}
\end{figure}

{\it Numerical simulation.} In the numerical simulation of the
driven  Van~der~Pol oscillator with piecewise-linear characteristic  
the dimensionless form of equations (\ref{eq3}) is used. 
The voltage-current characteristic of the nonlinear element 
(Fig.~\ref{fig2}) is transformed to dimensionless form also
(Fig.~\ref{fig5}). 
The analytical form of piecewise-linear $V$--$I$--characteristic 
could be described by equation
\begin{equation}
\label{eq4}
f(x) = {\left\{ {{\begin{array}{*{20}c}
 {9.09x + 10.09,} \hfill & {} \hfill & {x < - 1.0,} \hfill \\
 { - 1.0x,} \hfill & {} \hfill & { - 1.0 \le x \le 1.0,} \hfill \\
 {9.09x - 10.09,} \hfill & {} \hfill & {x > 1.} \hfill \\
\end{array}}}  \right.}
\end{equation}

Taking into account the values of the characteristic scales of the
piecewise linear nonlinearity ($I_{0} = 3.0$~mA, $V_{0} = 1.2$~V)
and the values
of the circuit elements (Fig.~\ref{fig1}) the dimensionless parameters 
($G=0.113$, $\varepsilon=1.75$) of the system (\ref{eq3}) are calculated.

\begin{figure}
\begin{center}
\leavevmode
\hbox{%
\epsfxsize=\columnwidth
\epsffile{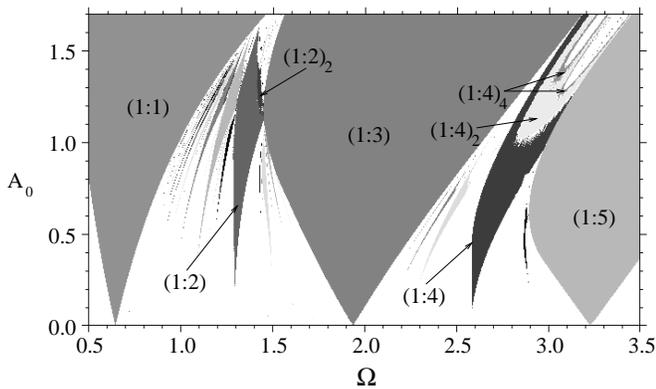}}
\end{center}
\caption{The parameter plane "the amplitude of external force -- the
frequency of external force" for the Van~der~Pol oscillator with
piecewise-linear $V$--$I$--characteristic of nonlinear element (numerical
simulation).}
\label{fig6}
\end{figure}

The parameter plane "the amplitude -- the frequency of
external force" for the system (\ref{eq3})
with piecewise-linear characteristic (\ref{eq4})
obtained in numerical simulation is shown in Fig.~\ref{fig6}, where 
synchronization regions are marked in gray scale. 
For plotting of the
parameter plane (Fig.~\ref{fig6}) the system (\ref{eq3}), (\ref{eq4})
was integrated by using of the
fourth order Runge-Kutta method with constant time step $h = 0.05$. The
identification of the synchronization regime was carried out from Poincare
section by selecting points in the  section through the period of the
external force. 

It is seen from Fig.~\ref{fig6}, that the characteristic 
synchronization tongues are obtained.  Main synchronization
tongues are marked, besides, many synchronization
tongues of higher order are also visible. The route to chaos via period
doublings is observed within the synchronization tongues
(1:2) and (1:4). Moreover, the view of the internal structure of the
synchronization tongues (1:2) and (1:4) allows to conclude 
that the {\it crossroad area} takes place
in a similar way as it was found in natural experiments with the 
electronic circuit
(Fig.~\ref{fig1}).

The visual comparison of the parameter planes obtained in the natural
experiment and the numerical simulation 
(Fig.~\ref{fig3} and \ref{fig6}, respectively) leads
to conclusion about qualitative agreement of both experiment and simulation.
The further comparisons could be done if to rescale the experimental results
(Fig.~\ref{fig3}) to numerical simulated ones 
(Fig.~\ref{fig6}) using the coefficients
introduced for obtaining the dimensionless equation (\ref{eq3}) 
then it is seen the
good quantitative agreement of both experiment and simulation.

{\it Conclusion.} 
In this paper the electronic model of the Van~der~Pol oscillator with
piecewise linear characteristic of nonlinear element is proposed. It is
discussed results of the experiment on the investigation of the
oscillator behaviour under external sinusoidal force 
at the parameter plane "the
amplitude -- the frequency of external force". The primary
attention has been paid to the revealing of regions on the parameter plane
in which the chaotic behaviour is realized and to investigations of internal
structure of such regions. It is shown that transition to chaos via period
doubling is taken place within the  synchronization tongues (1:2) and
(1:4) (here ratio of natural frequency of oscillator to external force
frequency) as well as in many synchronization tongues of higher order.

The comparison of the
parameter planes obtained in  experiment and numerical
simulation showed that the planes obey with both qualitative and
quantitative similarity for boundaries of regions with synchronous and
chaotic behaviour. 
To demonstrate this fact the typical points from 
the experimental parameter plane (Fig.~\ref{fig3}) could be rescaled to
dimensionless case and compared with the same points 
for the parameter plane obtained in numerical simulation
(Fig.~\ref{fig6}).  
So, for the point ($f_0=12.07$~kHz, $A'=0.55$~V) of minimum for $A'$  
at the boundary of (1:2) and (1:2)$_2$ regimes (Fig.~\ref{fig3}) could be
calculated dimensionless values ($\Omega=1.45$, $A_0=1.13$) that in good
agreement with the point of minimum for $A_0$ at the same boundary 
(Fig.~\ref{fig6}), analogously, for the point 
($f_0=25.72$~kHz, $A'=0.63$~V) of minimum at the boundary
(1:4)$_2$ and (1:4)$_4$ is obtained ($\Omega=3.08$, $A_0=1.30$).    
The region at the parameter plane where transition to
chaos takes place is displaced within the synchronization
tongues (1:2) and (1:4). The region has the {\it crossroad area} 
like view. The results of experimental investigations and numerical
simulations of the changes of the parameter plane structure 
for large values of external force and the circuit elements will be
published elsewhere.  

This work was partially supported by the RFBR under grant No.02--02--16351, 
the Ministry of Education of RF under grant No.E02--3.5--149
and the CRDF (BRHE REC--006).


\begin{thebibliography}{99}

\bibitem{r1}B.~Van~der~Pol and J.~Van~der~Mark, Nature {\bf 120}, 363 (1927).

\bibitem{r2}A.A.~Andronov, A.A.~Vitt, and S.E.~Khaikin, 
{\it Theory of Oscillators} (Pergamon Press, London, 1966).

\bibitem{r3}M.I.~Rabinovich and D.I.~Trubetskov, 
{\it Oscillations and Waves in Linear and
Nonlinear Systems} (Kluwer, Netherlands, 1989).

\bibitem{r3a}W.~Ebeling, {\it Structurbilding bei Irreversiblen
Prozessen} (Leipzig, 1976);
L.~Glass and M.C.~Mackey, {\it From Clocks to Chaos.
The Rhythms of Life} (Princeton University Press, Princeton, 1988);
H.G.~Schuster, {\it Deterministic Chaos: An Introduction } 
(VCH, Weinheim, 1988); 
W.-B.~Zhang, {\it Synergetic Economics} 
(Springer-Verlag, Berlin Heidelberg, 1991);
V.S.~Anishchenko, {\it Dynamical Chaos -- Models and Experiments}
(World Scientific, Singapore, 1995);
H.D.I.~Abarbanel, M.I.~Rabinovich, A.~Selverston, M.V.~Bazhenov, 
R.~Huerta, M.M.~Sushchik, and L.L.Rubchinskii,  
Phys. Usp. {\bf 39}, 337 (1996).

\bibitem{r3b}J.~Guckenheimer, IEEE Trans. 
Circuits  Syst. {\bf 27}, 983 (1980).

\bibitem{r4}M.P.~Kennedy and L.O.~Chua, IEEE Trans. 
Circuits Syst. {\bf 33}, 974 (1986). 

\bibitem{r5} Y.~Ueda and N.~Akamatsu, 
IEEE Trans. Circuits Syst. {\bf 28}, 217 (1981). 

\bibitem{r6} G.R.~Qin, D.C.~Gong, and X.D.~Wen, Phys. Lett. A 
{\bf 141}, 412 (1989). 

\bibitem{r7} U.~Parlitz and W.~Lauterborn, Phys. Rev. A 
{\bf 36}, 1428 (1987). 

\bibitem{r8} M.L.~Cartwright and J.E.~Litlewood, 
J. London Math. Soc. {\bf 20}, 180 (1945).

\bibitem{r9} N.~Levinson, Ann. Math. {\bf 50}, 127 (1949). 

\bibitem{r9a} R.~Mettin, U.~Parlitz, and W.~Lauterborn, 
Int. J. Bifurc. Chaos {\bf 3}, 1529 (1993).


\bibitem{r10} T.~Matsumoto, L.O.~Chua, and M.~Komuro, IEEE Trans. 
Circuits Syst. {\bf 32}, 798 (1985). 

\bibitem{r11} E.S.~Mchedlova, in {\it the Proceedings of the 5th
International Specialist Workshop on Nonlinear Dynamics of Electronic
Systems (NDES--97) (Moscow, Russia, June 26--27, 1997)} (Moscow, 1997), 
p.449--452.


\bibitem{r13} L.O.~Chua, C.~Desoer, and E.~Kuh, 
{\it Linear and Nonlinear Circuits} (McGrow--Hill
Book Company, 1987).

\bibitem{r14} J.P.~Carcasses, C.~Mira, M.~Bosch, C.~Simo, and J.C.~Tatjer,
Int. J. Bifurc. Chaos {\bf 1}, 183 (1991);
 A.P.~Kuznetsov, S.P.~Kuznetsov, and I.R.~Sataev,
Izvestiya VUZov. Prikladnaya Nelineynaya Dinamika. {\bf 1} (3/4), 17 (1993)
(in Russian).


\end{thebibliography}
\end{document}